\documentclass[11pt]{article}

\newcommand{\BC}{k_{\rm B}}

\usepackage{graphicx}
\usepackage{bm}

\begin{document}

\title{Modeling force-induced bio-polymer unfolding}
\author{Anthony J. Guttmann$^1$, Jesper L. Jacobsen$^2$, Iwan Jensen$^1$ 
and Sanjay Kumar$^3$ \\
$^1$ ARC Centre of Excellence for Mathematics and Statistics of Complex Systems, \\
Department of Mathematics and Statistics, \\
The University of Melbourne, Victoria 3010, Australia \\
$^2$ Universit\'e Paris Sud, UMR8626, LPTMS, F-91405 Orsay Cedex, France \\
   Service de Physique Th\'eorique, CEA Saclay, F-91191 Gif-sur-Yvette, France \\
$^3 $Department of Physics, Banaras Hindu University,
Varanasi 221 005, India}

\date{\today}

\maketitle

\begin{abstract}
We study the conformations of polymer chains in a poor solvent, with and without bending
rigidity, by means of a simple statistical mechanics model. This model can be exactly solved
for chains of length up to $N=55$ using exact enumeration techniques. We analyze in 
details the differences between the constant force and constant distance ensembles 
for large but finite $N$. At low temperatures, and in the constant force ensemble, 
the force-extension curve shows multiple plateaus (intermediate states), in contrast 
with the abrupt transition to an extended state prevailing in the $N \to \infty$ limit. 
In the constant distance ensemble, the same curve provides a unified response to pulling 
and compressing forces, and agrees qualitatively with recent experimental results. We 
identify a cross-over length, proportional to $N$, below which the
critical force of unfolding decreases with temperature, while above,
it increases with temperature. Finally, the force-extension curve for stiff chains 
exhibits ``saw-tooth" like behavior, as observed in protein unfolding experiments.
\end{abstract}


\section{Introduction}

In order to understand the role of molecular interactions in the
structural behaviour of bio-molecules, it has been found appropriate
to introduce force as a thermodynamic variable \cite{rief,busta}. This
development, which has occurred only in the last decade, permits the
experimental study of a variety of properties of bio-molecules,
including elastic, functional, mechanical and structural properties
\cite{busta1}. Furthermore, the variation of force with other parameters, such
as solvent pH, loading rate and temperature helps us understand the
interactions better \cite{evan,bloom,evans}. Many biological reactions involve large
conformational changes, with well defined responses. For example, the
radius of gyration of a polymer may display substantial variation, and
this can be used to follow the reaction as it progresses \cite{busta1}.

In these cases, a conceptually simple two-state model can be used to
track the process \cite{busta1}. The applied force twists the free energy
surface along the reaction co-ordinate by an amount proportional to
the radius of gyration. There are several important transitions of
this type, including the folding-unfolding transition \cite{rief}, the
stretching and unzipping of dsDNA \cite{bhat,bbs}, and the ball-string polymer
transition \cite{haupt}. Away from the $\theta$-temperature, we know that a
polymer will be in either a collapsed or a swollen state \cite{degennes}. The
mean-square radius of gyration $\langle R^2 \rangle_g$ scales with
chain length $N$ as $\langle R^2 \rangle_g \sim const. \times
N^{2\nu},$ where $\nu$ is the critical exponent. At low temperatures,
when the polymer is in the collapsed state, $\nu = 1/d$ while at high
temperature $\nu = 1, 3/4, 0.588\ldots, 1/2$ for $d=1,2,3,4$ \cite{degennes}
respectively. These values are believed to be exact for $d=1,2$ and,
with a logarithmic correction, for $d=4.$ For $d=3$ the (approximate)
numerical estimate is obtained from a variety of series expansion,
Monte Carlo and field theory methods, which are all in agreement
within error bars. It should be noted that a polymer cannot acquire
the conformation of a fully stretched state (characterised by $\nu =
1$), by varying the temperature alone. Application of sufficient force
will achieve this stretched state. Hence it can be seen that force not
only twists the free energy surface, but also introduces a new
stretched state, which is not otherwise accessible. Furthermore,
recent experiments \cite{haupt,hanbin,lemak} suggest that there are a number of
intermediate states achieved as the polymer is unfolded, which
manifest themselves as a sequence of plateaus in the force-scaled
elongation plots.

In studying such problems, there are two relevant ensembles. These are
the {\it constant force ensemble} ({\bf CFE}) and the {\it constant
distance ensemble} ({\bf CDE}). The former ensemble has predominantly
been used in the study of the non-equilibrium  thermodynamics of small
systems, where the average extension is taken as the control
parameter. In atomic force microscopy, the force is usually applied
using a linear ramp protocol with very small velocity, so changes take
place extremely slowly. In such cases the {\bf CDE} is the appropriate
ensemble. While one might expect the choice of ensemble to be
irrelevant in the thermodynamic limit \cite{busta2}, for small systems, such as
single molecules, the results are expected to be ensemble dependent
\cite{zema}. Other aspects of the experimental situation that have been
largely ignored include the loss of entropy due to the effective
confinement induced by such features as the attachment of receptor and
ligand molecules to a substrate and a transducer, respectively.

These transitions have typically been studied using very simple
models, such as the freely-jointed chain (FJC) or the worm-like chain
(WLC) \cite{fixman,doi}. These models ignore important effects such as excluded
volume, and attractive interactions, and though they have been used to
study the force-extension curves of bio-molecules \cite{doi}, they are
really only suited to do so in a good solvent. In a poor solvent, as
mentioned above, the force-extension curves display plateaus at
certain values of force \cite{haupt,hanbin}. A combination of an improved
theoretical understanding of semi-flexible polymers, combined with new
experimental observations of polymers in poor solvents have given rise
to improved levels of understanding of globules with a well defined
internal structure \cite{grass,garel}. These developments, in turn, have
potential application to the study and understanding of the basic
mechanism of protein folding.

In this paper we provide several results. Firstly a complete phase
diagram based on an analysis of exact enumerations of finite chains of
unprecedented lengths. We particularly focus our attention on the
low-temperature regime, for it is this region that is so difficult to
access either experimentally, or by other theoretical techniques, such
as Monte Carlo methods. These studies are very relevant to certain
features of single bio-molecules that are studied experimentally. In
particular, we find intermediate low-temperature states stabilised by
the force. We also find that there is a cross-over length, below which
the transition temperature is a decreasing function of force, while
above it increases with force.

We also introduce, for the first time, a bending rigidity, and repeat
our calculations with an additional parameter that tunes for
stiffness. For appropriate high values of bending rigidity, we find
saw-tooth like oscillations. These have been observed experimentally,
but as far as we are aware, have not been previously modelled.

A brief account of our work has appeared recently \cite{our_PRL}.
The present paper provides a more detailed discussion and several
new results.

In the next section we define our model precisely, and in section 3 we
calculate fluctuations in the number of contacts as a function of
force for fixed temperature, and vice versa. We use the positions of
the peaks in the fluctuation curves as a rough approximation to
determine points on the phase diagram, and use these to construct 
a qualitative `phase diagram'. In sections 4 and 5 we discuss the two
different ensembles, and discuss the effect of varying both
temperature and force. In section 6 we introduce a fugacity associated
with the number of bends in the polymer, and discover the new
phenomenon of a saw-tooth like behaviour in the force-extension curves
at sufficiently low temperatures. A brief summary and discussion
completes the paper.

\section{The model}
\begin{figure}
\begin{center}
\includegraphics[width=4.5in]{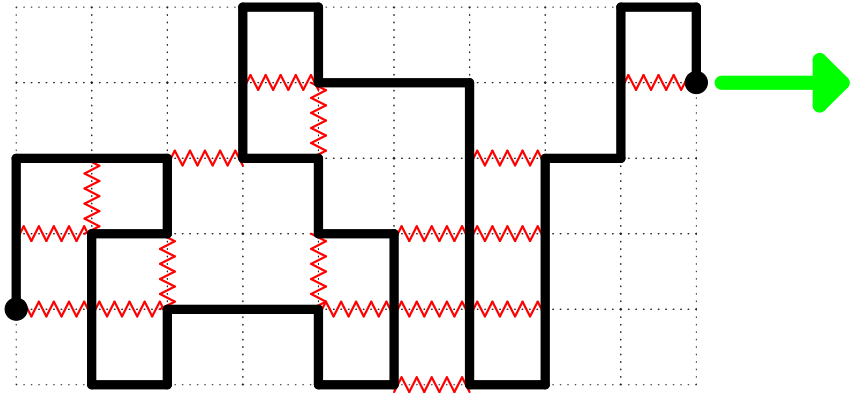}
\end{center}
\caption{\label{fig:model} An ISAW on the square lattice with one end
attached to a surface and subject to a pulling force on the other end.
Monomers are located on the vertices of the lattice and bonds are
indicated by the thick solid line. Interactions also occur between
non-bonded nearest neighbours as indicated by the zig-zag lines.}
\end{figure}

We model the polymer chains as interacting self-avoiding walks (ISAWs) 
on the square lattice \cite{vander} as shown in Fig.~\ref{fig:model}.
Interactions are introduced between non-bonded nearest neighbor monomers.
In our model one end of the polymer is attached to an  impenetrable neutral 
surface (there are no interactions with this surface) while the polymer 
is being pulled from the other end with a force acting along the 
$x$-axis. Note that the ISAW does not extend beyond either 
end-point so the $x$-coordinate $x_j$ of the $j$'th monomer is restricted
by $0=x_0 \leq x_j \leq x_N=x$. At first this restriction may appear
artificial but it does in fact model the experimental setup.
In typical experiments single proteins or small pieces of DNA 
are attached to the surface of beads (using ligand molecules). 
The beads are very large compared to the size of the proteins
and the surface of the bead is thus well approximated by a flat surface.

We introduce Boltzmann weights
$\omega=\exp(-\epsilon/\BC T$) and $u=\exp(-F/\BC T)$ conjugate
to the nearest neighbor interactions and force, respectively, where
$\epsilon$ is the interaction energy,
$\BC$ is Boltzmann's constant, $T$ the temperature and $F$ the
applied force. In the rest of this study we set $\epsilon=-1$ and $\BC=1$.
We study the finite-length partition functions
\begin{equation}
Z_N(F,T)  = \!\!\!\!\! \sum_{\rm all \ walks} \!\!\!\!\!\!\! \omega^m u^x 
     \! =  \sum_{m,x} \! C(N,m,x)  \omega^m u^x,
\end{equation}
where $C(N,m,x)$ is the number of ISAWs of length $N$ having $m$ nearest 
neighbor contacts and whose end-points are a distance $x=x_N-x_0$ apart.
The partition functions of the {\bf CFE}, $Z_N(F,T)$, and 
{\bf CDE}, $Z_N(x,T)= \sum_{m} C(N,m,x) \omega^m$, are 
related by $Z_N(F,T)  =  \sum_{x} Z_N(x,T) u^x$.
The free energies are evaluated from the partition functions 
\begin{eqnarray}
G(x)= -T \log Z_N(x) \; \; {\rm and } \; \;
G(F)= -T \log Z_N(F).
\end{eqnarray}
Here $\langle x \rangle = \frac{\partial G(F)}{\partial F}$ and 
$\langle F \rangle = \frac{\partial G(x)} {\partial x}$ are the 
control parameters of the {\bf CFE} and {\bf CDE},
respectively.

We enumerate all possible conformations of the ISAW by exact enumeration 
techniques. The major advantage of this approach is that the complete
finite-length partition functions can be analyzed exactly. 
Furthermore scaling corrections can be taken into account by suitable extrapolation 
schemes enabling us to obtain accurate estimates in the thermodynamic
(infinite length) limit \cite{gut}. To achieve a similar degree of accuracy using 
Monte Carlo simulations one typically has to use chains at least two orders 
of  magnitude longer than in the exact enumerations \cite{singh}.  
The greatest challenge facing exact enumerations is to increase the chain length. 
Until now most  exact results for models of small proteins were confined 
to chain lengths of 30 or so \cite{maren,kumar}. 
Here the number of  ISAWs was calculated using
transfer matrix techniques \cite{jensen}. Combined with parallel processing, we
were able to almost double the chain length. To be precise we calculate 
the partition functions up to chain length 55.

\section{Fluctuation curves and phase diagram}

\begin{figure}
\begin{center}
\includegraphics[width=4.5in]{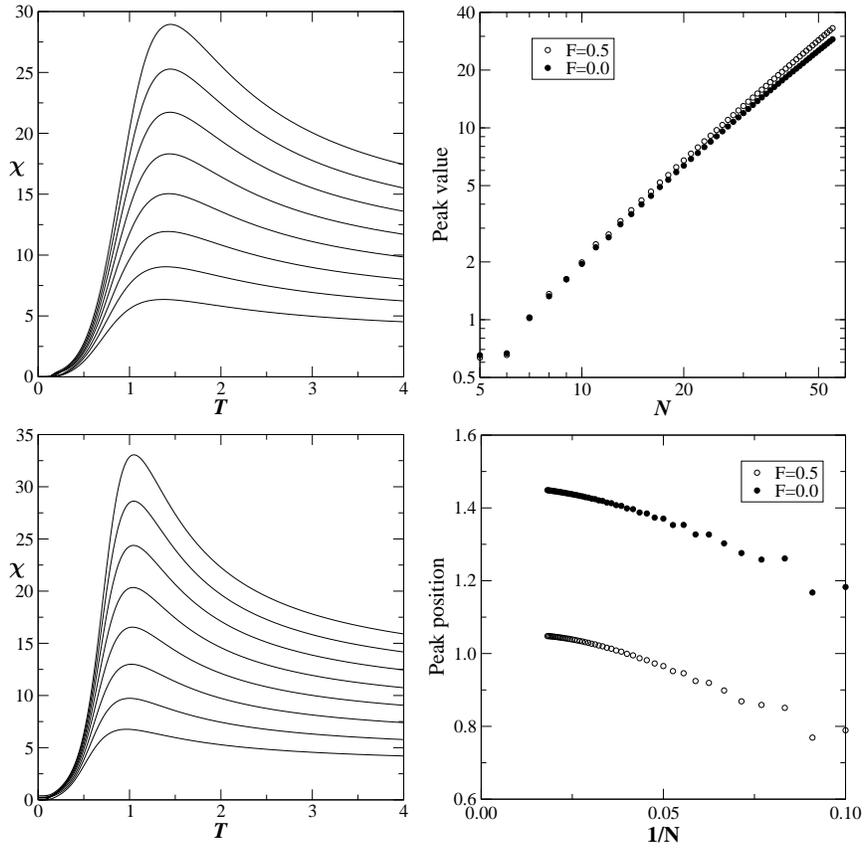}
\end{center}
\caption{\label{fig:FlucF} The fluctuations in the number of contacts
as a function of temperature for fixed force $F=0.0$ (upper left panel)
and $F=0.5$ (lower left panel). Each panel contains curves for ISAWs of
length (from bottom to top) $N=20, 25,\ldots, 55$. In the upper right panel
we show a log-log plot of the growth in the peak value of the fluctuation curve
with chain length $N$. The lower right panel shows the peak position (critical 
temperature value) vs $1/N$. 
}
\end{figure}

At low temperature and force the polymer
chain is in the collapsed state and as the temperature is increased (at fixed
force) the polymer chain undergoes a phase transition to an extended state.
The value of the transition temperature (for a fixed value of the force) can
be obtained  from the fluctuations in the number of
non-bonded nearest neighbor contacts (which we can calculate exactly for
any finite $N$ up to 55).
The fluctuations are defined as $\chi=\langle m^2 \rangle -\langle m \rangle^2$,
with the $k$'th moment given by 
$$\langle m^k \rangle = 
\frac{\sum_{m,x} \! m^k C(N,m,x)  \omega^m u^x}{\sum_{m,x} \! C(N,m,x)  \omega^m u^x}.$$
In the panels of Fig.~\ref{fig:FlucF} we show the emergence of peaks in 
the fluctuation curves with increasing $N$ at fixed force $F=0.0$ and $F=0.5$. 
In the top right panel we show the growth in the peak value as $N$ is increased.
Since this is a log-log plot we see that the peak values grows as a power-law with
increasing $N$; this divergence is the hall-mark of a phase transition. In the lower
right panel we have plotted the position of the peak (or transition temperature)
as a functions of $1/N$. Clearly the transition temperature appears to converge to
a finite (non-zero) value but the data exhibits clear curvature which makes an
interpolation to infinite length difficult.

\begin{figure}
\begin{center}
\includegraphics[width=4.5in]{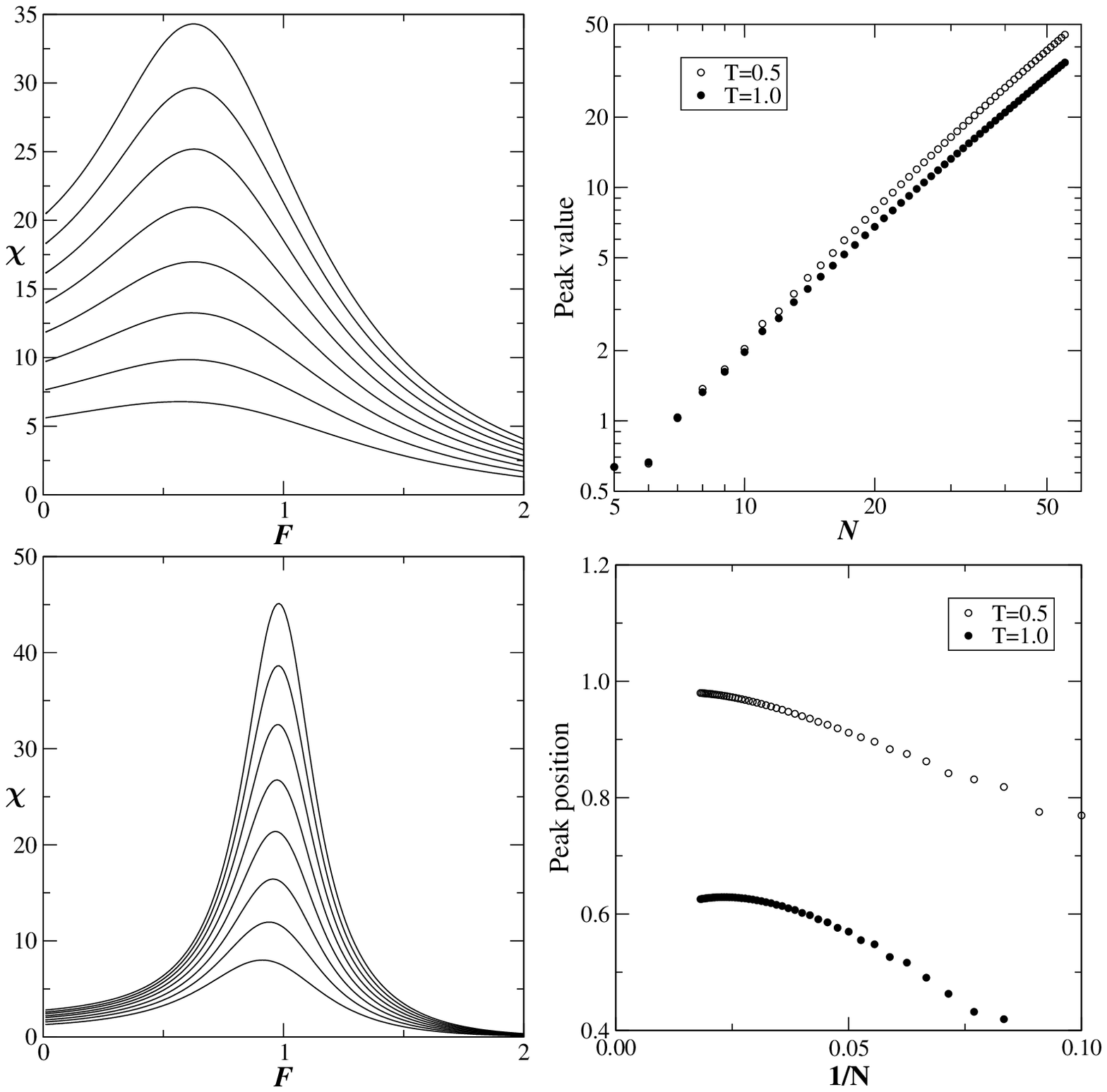}
\end{center}
\caption{\label{fig:FlucT} The fluctuations in the number of contacts
as a function of force for fixed temperature $T=1.0$ (upper left panel)
and $T=0.5$ (lower left panel). Each panel contains curves for ISAWs of
length (from bottom to top) $N=20, 25,\ldots, 55$. In the upper right panel
we show a log-log plot of the growth in the peak value of the fluctuation curve
with chain length $N$. The lower right panel shows the peak position (critical 
force value) vs $1/N$. }
\end{figure}

One can also study the same transition phenomenon by fixing the temperature
and varying the force. In the panels of Fig.~\ref{fig:FlucT} we show the emergence 
of peaks in  the fluctuation curves with increasing $N$ at fixed temperature
$T=1.0$ and $T=0.5$. Again we observe the power-law divergence of the peak-value.
The only other note-worthy feature is that in the plots of the peak position
(critical force value) we observe not only strong curvature but we actually
see a turning point in the curves as $N$ is increased. This feature would make it
impossible (given the currently available chain lengths) to extrapolate this data.

\begin{figure}
\begin{center}
\includegraphics[width=4.5in]{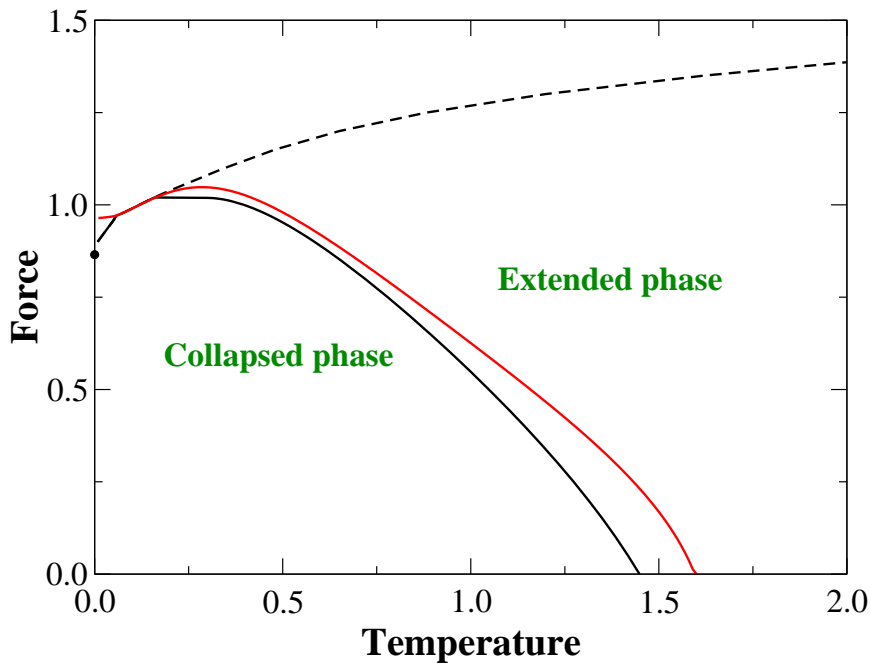}
\end{center}
\caption{\label{fig:phase} The `phase diagram'  for flexible chains as obtained
from the position of the peak in the contact fluctuation curves for $N=55$. The solid
black curve and the dashed curve are obtained by fixing the force and varying 
the temperature.
The solid red curve is obtained by fixing the temperature and varying the force.}
\end{figure}

\begin{figure}
\begin{center}
\includegraphics[width=4.5in]{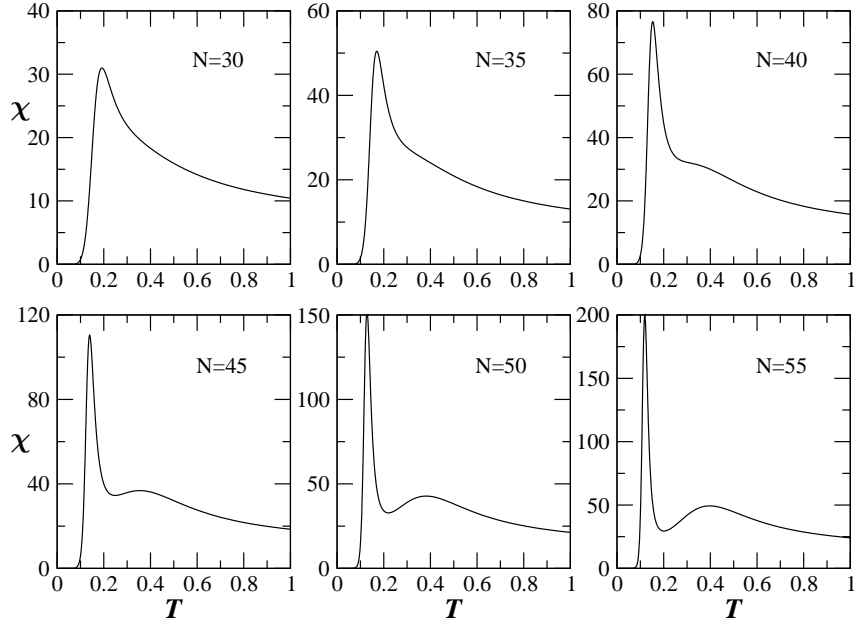}
\end{center}
\caption{\label{fig:FlucF100} The fluctuations in the number of contacts
as a function of temperature for fixed force $F=1.0$.}
\end{figure}

In Fig.~\ref{fig:phase}, we show the force-temperature `phase diagram'
for flexible chains as obtained from the peak positions (note that the
true phase diagram should be obtained by extrapolating our data to
the $N\to \infty$ limit). 
The qualitative features of the phase diagram remains 
largely the same as those observed in previous studies \cite{kumar}. 
In Fig.~\ref{fig:phase} we have shown the transitions as obtained
by fixing the force (black curves) or fixing the temperature (red curve).
One of the most notable feature of  the phase-diagram is the
{\em re-entrant} behaviour which can be explained 
by a ground state with non-zero entropy \cite{kumar,ret}. 
Using phenomenological
arguments at $T=0$, the critical force  $F_c=0.8651$ (indicated by a black circle
on the $y$-axis) is found from the expression,
$F_c =  1 - 1/\sqrt{N}+  T S$ \cite{kumar}.
The positive slope $dF_c/dT$ at $T=0$ confirms the existence of re-entrance
in the $F-T$ phase-diagram.
In the panels of Fig.~\ref{fig:FlucF100} we show the emergence of two peaks in 
the fluctuation curves with increasing $N$ at fixed force $F=1.0$. The twin-peaks
reflect the fact that in the re-entrant region as we increase $T$ (with $F$ fixed)
the polymer chain undergoes two phase transitions. Note that the twin-peaks are
not apparent for small values of $N$.

\begin{figure}
\begin{center}
\includegraphics[width=4.5in]{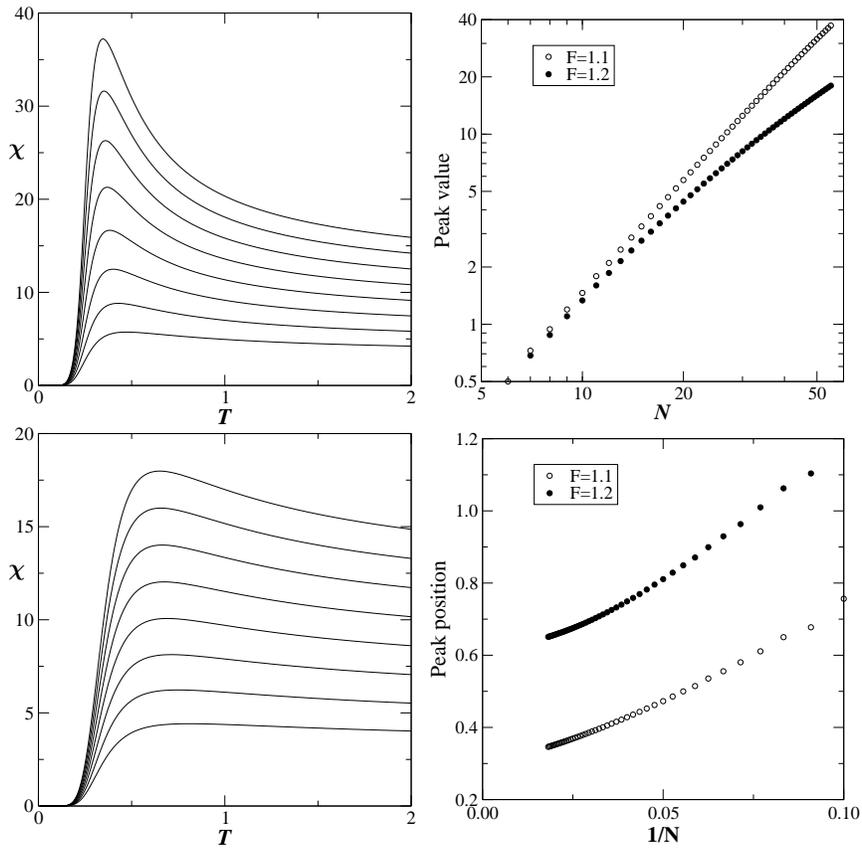}
\end{center}
\caption{\label{fig:FlucHighF} The fluctuations in the number of contacts
as a function of temperature for fixed force $F=1.1$ (upper left panel)
and $F=1.2$ (lower left panel). Each panel contains curves for ISAWs of
length (from bottom to top) $N=20, 25,\ldots, 55$. In the upper right panel
we show a log-log plot of the growth in the peak value of the fluctuation curve
with chain length $N$. The lower right panel shows the peak position (critical 
temperature value) vs $1/N$. 
}
\end{figure}

The other notable feature is that in the fixed force case we see a new transition line 
from the extended state to the fully stretched
state which is solely induced by the applied force (the dashed line
in Fig.~\ref{fig:phase}. 
In contrast to the
lower phase boundary (collapse transition), where the force decreases 
with temperature, the upper phase boundary (stretching transition) 
shows that the force increases with the temperature. 
However, very curiously, we {\em do not} observe this upper transition line
when we fix the temperature and let the force vary. Indeed this is clear
from  Fig.~\ref{fig:FlucT} where at fixed $T=0.5$ and $1.0$ we see only
a single peak (giving us points on the red curve in the phase diagram 
Fig.~\ref{fig:phase}). In Fig.~\ref{fig:FlucT} the value of the force extends
up to $F=2.0$ and the upper transition (dashed line in the phase diagram)
should appear (if present) as a second peak in the fluctuation curves
of Fig.~\ref{fig:FlucT}. The absence of any evidence of a second peak
is what leads us conclude that we do not see this second transition 
in the fixed $T$ varying $F$ study. We are not sure whether the 
upper phase boundary is a real phase transition or a crossover phenomenon. 
If it is a real phase transition then the phase-diagram does not
really exhibit re-entrance. 
In re-entrance we go from phase A to phase B and then again to phase A.
In Fig.~\ref{fig:FlucHighF} we have plotted
the fluctuation curves for force $F=1.1$ and $F=1.2$. The curves for
$F=1.1$ (including the plot of the peak height) looks very similar to
the plots (see Fig.~\ref{fig:FlucT}) for low values of the force. For
force $F=1.2$ the peak is not very pronounced and we are hesitant to
even call it a peak. Also when we look at the peak height vs. $N$
it appears that the curve has two different behaviours for small
and large $N$, respectively. This could be a sign of a cross-over behaviour. 
We are currently performing
more studies to try to understand and resolve this discrepancy.

\section{The constant force ensemble}

\begin{figure}
\begin{center}
\includegraphics[width=4.5in]{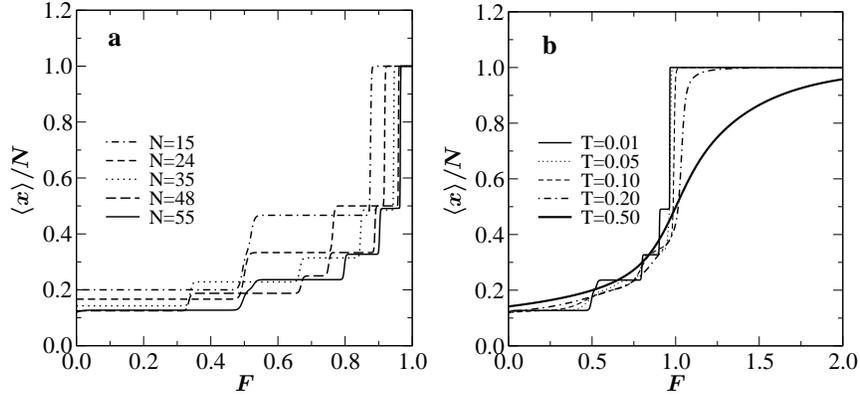}
\end{center}
\caption{ \label{fig:AveX}
The average scaled elongation $\langle x \rangle/N$ vs $F$ at $T=0.01$ 
for various lengths (a) and temperatures at length $N=55$ (b).}
\end{figure}

First we study the model in the {\bf CFE}.
In Fig.~\ref{fig:AveX}a, we plot the average scaled elongation,
$$\langle x \rangle/N=
\frac{1}{N}\frac{\sum_{m,x} \! x C(N,m,x) \omega^m u^x}{\sum_{m,x} \! C(N,m,x)  \omega^m u^x},$$
as a function of the applied force 
for different chain lengths at low temperature. Experimentally several
transitions were found  in the force-extension curves corresponding to
many intermediate states \cite{haupt,hanbin,lemak,rief2}. This phenomenon is clearly
confirmed by our study. It has been argued that in the limit of infinite 
chain length, the intermediate states should  vanish  and there will be 
an abrupt transition between a folded and fully extended state \cite{maren}. 
Evidence to this effect can also be seen in Fig.~\ref{fig:AveX}a where we note
that the plateaus at an extension around $0.2$ tends to increase with $N$
while the other plateaus tend to shrink with $N$ (this is particularly so
for the plateau around $0.5$ corresponding to a simple zig-zag pattern of the chain).
Note however, that as we change the chain length from $25$ to $55$, we find 
more and more of these intermediate states. This has also been observed in recent
experiments \cite{haupt,rief2} where the globule deforms into an ellipse and then into a
cylinder. At a critical extension the polymer undergoes a sharp
first order transition into a ``ball string" conformation \cite{haupt,rief2}. This 
shows that finite size effects are crucial in all the single 
molecule experiments and can be seen even for long chains \cite{lemak}.

A simple theoretical argument for the observed behavior is that at low temperature,
where the entropy $S$ (per monomer) of the chain is quite low, the dominant 
contribution to the free energy
\begin{equation}
G =  N \epsilon - \sigma(N,F) \epsilon - N T S
\end {equation}
is the non-bonded nearest-neighbor interaction $N\epsilon$.
The second term is a surface correction and it vanishes in the
thermodynamic limit. However for finite $N$, the system has many
degenerate states depending upon the shape of the
globule. This leads to a surface correction term  $\sigma(N,F)$
which is a function of $N$ and $F$. If $F=0$ the
shape of the globule is like a square and the surface
correction term $\sigma(N,0)$ will be minimized and
equal to $2 \sqrt{N}$. In the {\bf CFE}, there is a
force induced additional contribution proportional to
the extension of the globule which along with $\sigma(N,F)$
stabilizes the intermediate states. When the temperature increases the 
multi-step character of the force-extension curve is  washed out due to
increased contributions from entropy. This effect can be clearly seen in
Fig.~\ref{fig:AveX}b  where we have plotted force-extension curves at 
different $T$.

\begin{figure}
\begin{center}
\includegraphics[width=3.0in]{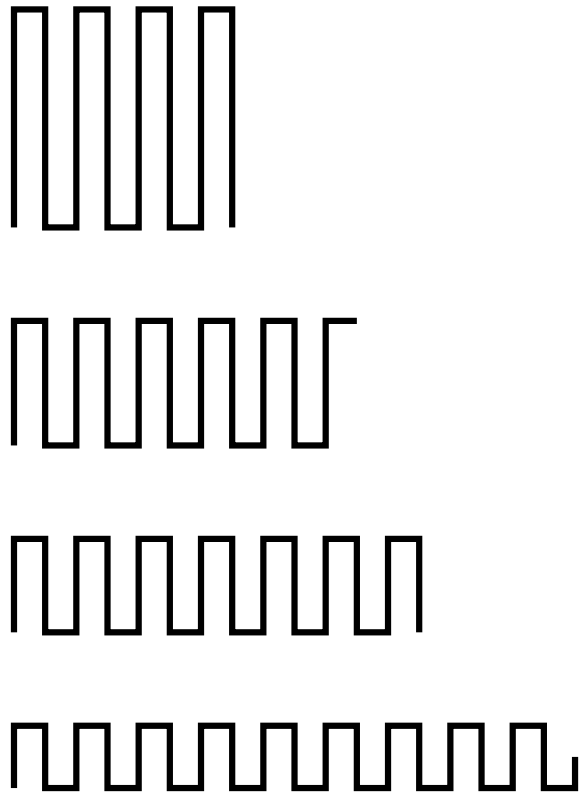}
\end{center}
\caption{ \label{fig:zigzag}
Conformations responsible for the plateaus in the force-elongation curve
with $T \to 0$ and $N=55$.}
\end{figure}

In the limit $T \to 0$, and for any fixed $N$, we can make the argument more precise,
and deduce the exact values of the elongation at the plateaus, and the exact values
of the force at the transition between two plateaus. Consider for instance the case $N=55$,
cf. Fig. \ref{fig:AveX}a. Our task is to find the conformations that minimize the energy,
i.e., maximize the quantity $F x + m$. For $F \to 0$ this means just maximizing $m$, and
this is achieved by zigzag conformations inscribed in a rectangle which is as close to a square
as possible. The best choice (unique up to trivial lattice symmetries) is the 
$x \times y$ = $7 \times 6$ rectangle shown in Fig.~\ref{fig:zigzag}a, which has $m=42$, 
and is responsible for the plateau at $\frac{x}{N} = \frac{7}{55} \simeq 0.127$. From this 
conformation it is easy to find other with $m \to m-1$ and $x \to x+1$ by unzipping a monomer 
at the boundary, but this is not favorable for $F<1$. However, at $F=\frac{1}{2}$ the 
$11 \times 4$ conformation of Fig.~\ref{fig:zigzag}b with $m=40$, and the $13 \times 3$ 
conformation of Fig.~\ref{fig:zigzag}c with $m=39$ start participating. The former is 
responsible for the shoulder at $\frac{x}{N}=\frac{11}{55} = 0.200$, clearly visible 
in Fig.~\ref{fig:AveX}a. However, the latter becomes the ground state for $F>\frac{1}{2}$, 
and so the next plateau is at $\frac{x}{N}=\frac{13}{55} \simeq 0.236$. Finally, the 
$18 \times 2$ conformation of  Fig.~\ref{fig:zigzag}d with $m=35$ starts dominating 
at $F=\frac{4}{5}$ and is responsible for the plateau at $\frac{x}{N} = \frac{18}{55} \simeq 0.327$.

It is thus evident that the precise plateau structure is extremely dependent on the value 
of $N$ and the details of the model (e.g., the choice of lattice). However, the following 
intuitive picture is of a more general validity: The dominant conformation for $F=0$ fills 
out a square, and as $F$ is increased more elongated rectangular conformations start dominating. 
Each of the plateau transitions corresponds to making the rectangle one or a few units thinner 
in the $y$ direction. In the large $N$ limit, a direct comparison of the conformations with 
$y=\sqrt{N}$ and $y=\sqrt{N}-1$ shows that indeed the first plateau will dominate for all 
$F \in [0,1)$, as expected from an inspection of Fig.~\ref{fig:AveX}a,
and in agreement with Ref.~\cite{maren}.

\section{The constant distance ensemble}

\begin{figure}
\begin{center}
\includegraphics[width=4.5in]{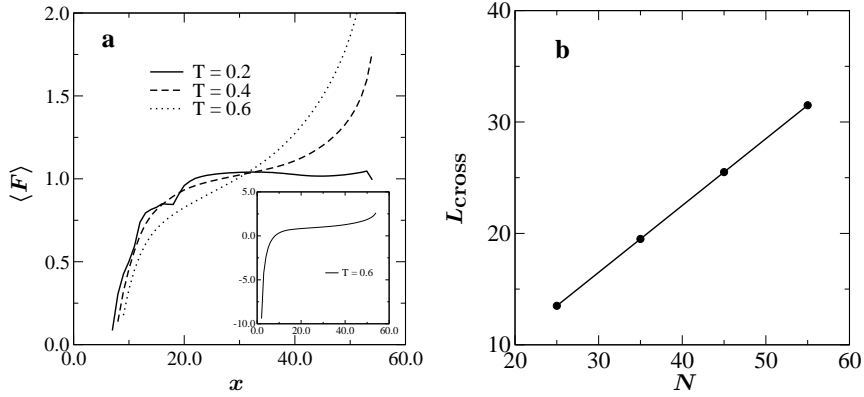}
\end{center}
\caption{ \label{fig:AveF} 
Plot of the average force $\langle F \rangle$ vs the elongation $x$ at 
various temperatures $T$ for $N =55$ (a) and the cross-over length vs $N$ (b).}
\end{figure}

Next we study the model in the {\bf CDE}.
The force-extension curve shown in the insert of 
Fig.~\ref{fig:AveF}a has interesting 
features. It shows that when the distance between the first
and the last monomer (where force is applied) is less than 
the average size of the coil (without force), one needs a
compressing force instead of a pulling force.
The qualitative behavior is similar to one observed in experiments \cite{sevick}
and computer simulations \cite{jorge}.
Since most models do not include confinement in their description, 
such behavior could not be predicted.  
In Fig.~\ref{fig:AveF}a, we show the response of the force when the elongation exceeds
the average size of the polymer. The flat portion of the 
curve gives the average force needed to unfold the chain.
Such plateaus have been 
seen in experiments \cite{haupt,lemak,hanbin}. From  Fig.~\ref{fig:AveF}a one 
can also see  that
the force required to obtain a given extension initially decreases with temperature. 
But beyond a certain extension (close to 30 in this case) the required force
increases with temperature. We note that the curves 
cross each other at a `critical' extension for any temperature (below
the $\theta$-point). We identify this as a cross-over point. In 
Fig.~\ref{fig:AveF}b we plot the position of the cross-over point 
$L_{cross}$ as a function of the length $N$ of the polymer chain. 
We see that the crossover extension increases linearly with the chain length.
This shows that above this point the chain acquires the conformation of the 
stretched state.
The increase in force with temperature generates a tension 
in the chain sufficient to overcome the entropic effect. Since the contribution to 
the free energy from this term is $TS$ ($S$ being the entropy), more force 
is needed at higher $T$ as seen in experiments. Our exact analysis for 
finite chain length shows that applying a force at first favors 
taking the polymer from the folded state to the unfolded state. However, 
rupture or second unfolding occurs when the tethered or unfolded chain attains the 
stretched state, and one requires more force at higher temperature.

\section{Semi-flexible polymers}

We model semi-flexible polymers by associating a positive energy 
$\epsilon_b$ with each turn or bend of the walk \cite{kumar}. 
The corresponding Boltzmann 
weight is $\omega_b=\exp(-b\epsilon_b)$, where $b$ is the number of bends
in the ISAW. We again enumerate all walks, but because of
the additional parameter $\omega_b$, we were restricted to $45$ steps.
For a semi-flexible polymer chain, a stretched state may be favored by 
increasing the stiffness. The phase diagram for semi-flexible 
chains is now well established. It has three states namely
(i) an open coil state at high temperature, (ii) a molten globule
at low temperature and low stiffness and (iii) a 'frozen' or 'folded'
state at low temperature and large stiffness \cite{grass,garel,kumar}. 
We note that while the flexible and semi-flexible $F-T$ phase-diagrams 
are qualitatively similar \cite{kumar}, the re-entrant behavior is suppressed
because of stiffness and becomes less pronounced with
increasing bending energy.

\begin{figure}
\begin{center}
\includegraphics[width=4.5in]{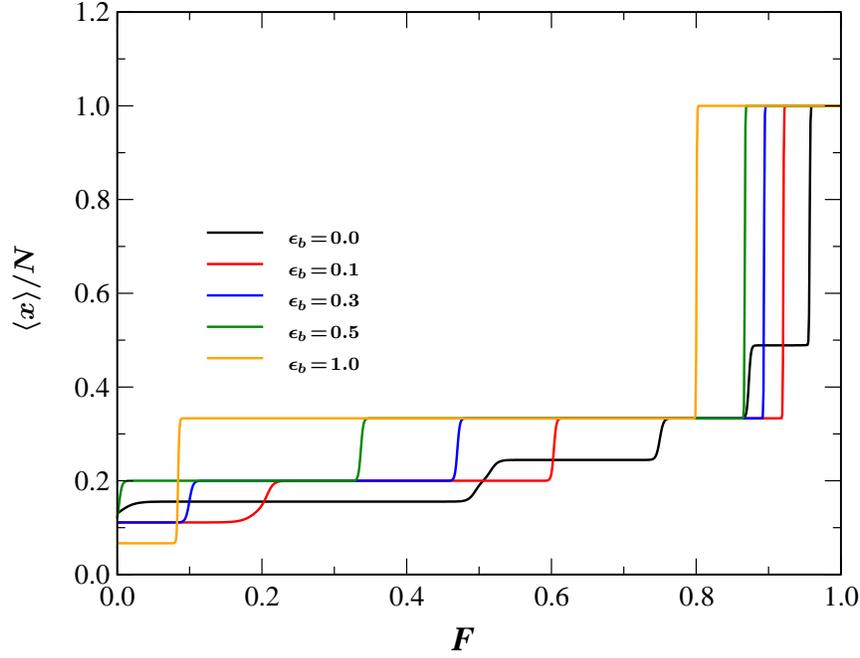}
\end{center}
\caption{\label{fig:AveX-semi} Plot of the average scaled
elongation $\langle x \rangle$/N vs $F$ 
for  semi-flexible chains with bending energy ranging from $\epsilon_b=0.0$
(flexible chains) to $\epsilon_b=1.0$ at low temperature $T=0.01$ for 
chain length $N =45$}
\end{figure}

In Fig.~\ref{fig:AveX-semi}, we plot the force-extension curves
for semi-flexible chains  with bending energy ranging from $\epsilon_b=0.0$
(fully flexible chains) to $\epsilon_b=1.0$ at low temperature $T=0.01$ for 
chain length $N =45$. We observe that as the bending energy is increased
the chain undergoes fewer intermediate transitions between the compact
state (low force) and the fully stretched state (high force). In particular
the flexible chain ($\epsilon_b=0.0$) has five plateaus while the chain
with $\epsilon_b=1.0$ has only three. This can again be explained more
quantitatively by inspecting the dominant configurations in the limit
$T\to 0$, as in section 4 above.

\begin{figure}
\begin{center}
\includegraphics[width=6in]{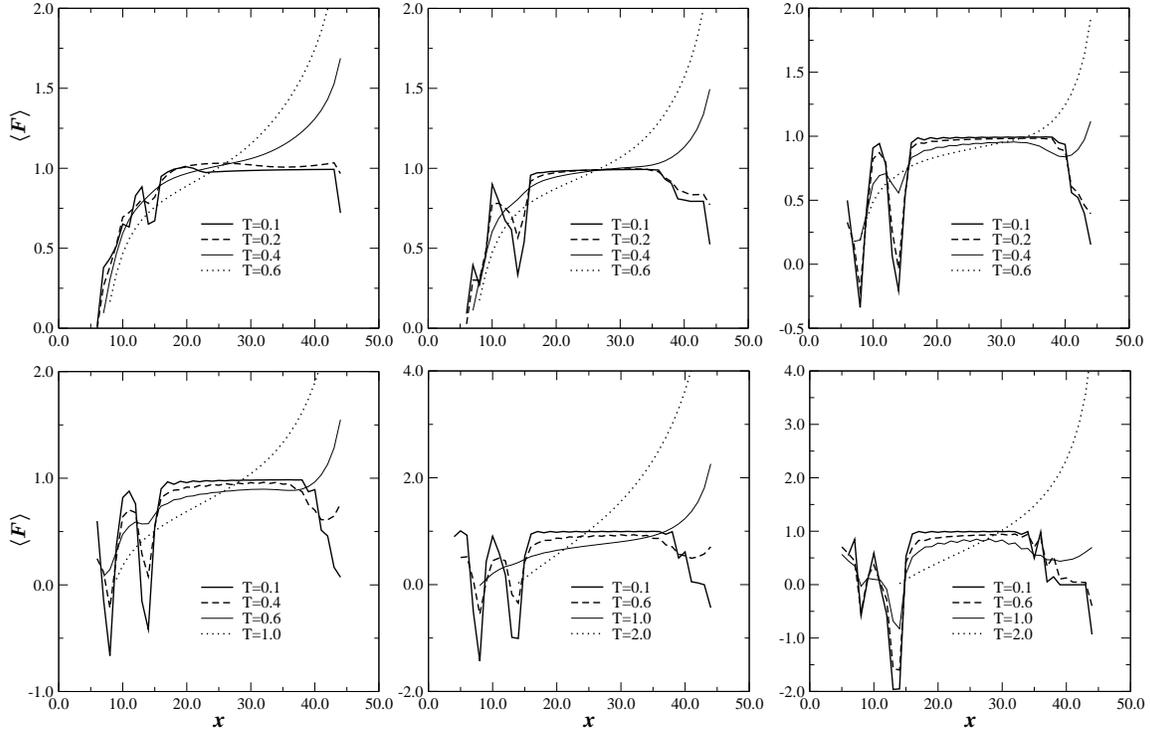}
\end{center}
\caption{\label{fig:AveF-semi} Plot of $\langle F \rangle$ vs $x$ 
for a semi-flexible chain with bending energy (from left to right and top to bottom)
$\epsilon_b=0.0,\, 0.1,\, 0.3,\, 0.5,\, 1.0$ and $2.0$
at different temperatures $T$ for $N =45$}
\end{figure}

In the {\bf CFE}, the probability distribution of the end-to-end
distance has "saw-tooth" like behavior corresponding to
intermediate states during unfolding \cite{kumar}. Therefore, it is important
to study the effect of stiffness on force-extension curves in the 
{\bf CDE}. 
The force extension curves shown in Fig.~\ref{fig:AveF-semi} have striking 
differences from the flexible ones. At low temperatures we see
strong oscillations which vanish as the temperature is increased. 
As the bending energy is increased and the polymer becomes more rigid
the oscillatory behavior extends to higher and higher temperatures.
Since the polymer chain has ``frozen conformations" like $\beta-$sheets 
(the zero-force limit of which describes zig-zag configurations inscribed 
in a square), it takes more force to unfold a layer. When about half a layer
has been opened, the bending energy favors a complete stretching of
the layers and hence the force decreases. This phenomenon
allows us to probe a molecule like Titin which has similar 
$\beta$-sheet structure \cite{rief1}.

\section{Summary and discussion}

To summarize, we have presented the exact solution of a model of long (finite)
polymer chains, of direct relevance to recent experiments on the elastic
properties of single biomolecules. The model takes into account several
constraints imposed by the experimental setups: geometric constraints,
excluded volume effects,
attraction between chain segments, finite but large chain length (here up
to $N=55$). It permits one to choose the thermodynamic ensemble ({\bf CFE} 
or {\bf CDE}) dictated by the experimental protocol.
The exact enumeration data permits us to access all parameter values, including
biologically relevant low temperatures where previous studies have failed.
Our results correctly reproduce several experimentally observed effects:
multiple transitions during unfolding, ``saw-tooth'' like oscillations in the
force-extension curve of semi-flexible chains and first-order transition into
a ``ball string'' conformation. Finally, we have identified cross-over
behavior that provides a unified treatment of both pulling and compressing forces 
in the {\bf CDE}.

\section*{Acknowledgments}

We would like to thank
the Australian 
Research Council (IJ,AJG), the Indo-French Centre for the Promotion 
of Advanced Research (CEFIPRA) (JLJ) and  MPIPKS, Dresden, Germany (SK).
The calculations presented in this paper used the computational resources of the
Australian Partnership for Advanced Computing (APAC) and the  
Victorian Partnership for Advanced Computing (VPAC).

\end{document}